%%%%%%%%%%%%%%%%%%%%%%%%%%%%%%%%%%%%%%%%%%%%%%%%%%%%%%%%%%%%%%%%%%%%%%%%%%%%%%%
% Comment on ``Comparison of approaches to classical signature change''
%%%%%%%%%%%%%%%%%%%%%%%%%%%%%%%%%%%%%%%%%%%%%%%%%%%%%%%%%%%%%%%%%%%%%%%%%%%%%%%

\def\bs{\bigskip}
\def\ms{\medskip}
\def\np{\vfill\eject}

\def\ni{\noindent}
\def\cl{\centerline}

\def\ref#1#2#3#4{#1\ {\it#2\ }{\bf#3\ }#4\par}
\def\refb#1#2#3{#1\ {\it#2\ }#3\par}
\def\PR{Phys.\ Rev.}

\def\l{\lambda}
\def\d{\delta}
\def\e{\varepsilon}
\def\p{\partial}
\def\half{{\textstyle{1\over2}}}
\def\implies{\quad\Rightarrow\quad}

\magnification=\magstep1
\overfullrule=0pt

\cl{\bf Comment on ``Comparison of approaches to classical signature change''}
\bs\cl{\bf Sean A. Hayward}
\cl{Department of Physics,
Kyoto University,
Kyoto 606-01,
Japan}
\bs\ni
{\sl Abstract.}
This is a comment on a reply to a comment on a paper of Hellaby \& Dray,
repeating the identification of an important mistake 
which is still being denied by the authors:
their proposed solutions do not satisfy 
the Einstein-Klein-Gordon equations at a change of signature.
Substitution of the proposed solutions into the Einstein-Klein-Gordon equations
in unit normal coordinates yields Dirac $\delta$ terms
describing source layers at the junction.
Hellaby \& Dray's criticisms of this straightforward calculation are absurd:
it does not involve ``imaginary time'',
it does not involve a ``modified form'' of the field equations,
and it is ``purely classical''.
Moreover, Hellaby \& Dray's latest attempt to lose the $\delta$ terms 
is mathematically invalid, 
involving division by zero and products of distributions,
hinging on an identity whose incorrectness may be checked by substitution.
\bs\ni
{\sl I. Introduction}

Hellaby \& Dray [1] claimed that matter conservation fails 
at a change of signature,
based on earlier work reporting invalid solutions 
to the Einstein or Klein-Gordon equations which I had already corrected.
I pointed out the error again in a comment [2] on [1],
but Hellaby \& Dray again denied it in a reply [3].
This is another comment on their reply.
Other relevant references may be found in [1--3].

Firstly, the calculation of the field equations given in [2] will be repeated,
specifically pointing out which components they represent 
and substituting the proposed solutions of Hellaby \& Dray explicitly,
showing that they fail to satisfy the field equations 
by certain well defined Dirac $\d$ terms 
describing source layers at the junction.
Secondly, 
Hellaby \& Dray's main criticisms [3] of [2] will be rebutted briefly;
they are all obviously false.
Thirdly, Hellaby \& Dray's latest attempt [3] 
to justify their proposed solutions will be shown to be mathematically invalid.
\ms\ni
{\sl II. The Einstein-Klein-Gordon equations}

The example considered in [2] was homogeneous isotropic cosmologies, given by
$$ds^2=-\l^{-1}dt^2+a^2\left({dr^2\over{1-kr^2}}+r^2d\Omega^2\right)\eqno(1)$$
where $d\Omega^2$ refers to a unit sphere, $k\in\{-1,0,1\}$
and the scale factor $a$ and inverse squared lapse $\l$ are functions of $t$.
Here the notation of [2], using $N$ for $\l$,
has been changed to avoid confusion with [3], which uses $N=\l^{-1}$.
The real scalar field $\phi$ is a function of $t$, with potential $V(\phi)$.
A straightforward formal calculation yields
$$\eqalignno{
&G^t_t=-3a^{-2}\left(k+\l(a')^2\right)&(2a)\cr
&G^r_r=-a^{-1}\left(\l'a'+2\l a''\right)-a^{-2}\left(k+\l(a')^2\right)&(2b)\cr
&T^t_t=-\half\l(\phi')^2-V&(2c)\cr
&T^r_r=\half\l(\phi')^2-V&(2d)\cr
&\nabla^2\phi=-\half\l'\phi'-\l\phi''-3\l a^{-1}a'\phi'&(2e)\cr}$$
where $G^i_j$ is the mixed Einstein tensor,
$T^i_j$ is the mixed stress-energy tensor,
$\nabla^2$ is the d'Alembertian 
and the prime denotes $\p/\p t$.
The independent components of the Einstein-Klein-Gordon equations are
$$\eqalignno{
&G^t_t=\kappa T^t_t&(3a)\cr
&G^r_r=\kappa T^r_r&(3b)\cr
&\nabla^2\phi=\p V/\p\phi&(3c)\cr}$$
where $\kappa=8\pi G$.
These field equations are linear combinations of the equations given 
with units $16\pi G=1$ in my previous comment [2].
They also agree formally with the equations (2)--(3) of Hellaby \& Dray [3]
apart from a curious factor in $V$.

Taking $\l=\e$, where $\e=\hbox{sign}(t)$,
describes a change of signature of the contravariant metric 
in unit normal coordinates,
i.e.\ positive $t$ is proper time and negative $t$ 
is proper distance from the junction $t=0$.
For clarity, denote this choice of $t$ by $\tau$
and the corresponding derivative by a dot rather than a prime.
Since $\dot\e=2\d$,
where $\d$ is the Dirac distribution in $\tau$ 
with support at the junction $\tau=0$,
direct substitution yields
$$\eqalignno{
&G^\tau_\tau-\kappa T^\tau_\tau=-3a^{-2}\left(k+\e\dot a^2\right)
+\kappa\left(\half\e\dot\phi^2+V\right)&(4a)\cr
&G^r_r-\kappa T^r_r=-2a^{-1}\left(\d\dot a+\e\ddot a\right)
-a^{-2}\left(k+\e\dot a^2\right)
+\kappa\left(V-\half\e\dot\phi^2\right)&(4b)\cr
&\nabla^2\phi-\p V/\p\phi
=-\d\dot\phi-\e\ddot\phi-3\e a^{-1}\dot a\dot\phi-\p V/\p\phi.&(4c)\cr}$$
The field equations are evidently well defined 
as distributional equations in the sense of Schwartz 
if $a$ and $\phi$ are differentiable at $\tau=0$ 
and twice differentiable for $\tau\not=0$,
and $V$ is differentiable.
Due to the terms in $\d$, the field equations require the junction conditions:
$$\eqalignno{
&G^i_j=\kappa T^i_j\implies\dot a\d=0\implies\dot a|_{\tau=0}=0&(5a)\cr
&\nabla^2\phi=\p V/\p\phi\implies\dot\phi\d=0
\implies\dot\phi|_{\tau=0}=0.&(5b)\cr}$$
The proposed solutions of Hellaby \& Dray do not satisfy these conditions,
but only the remaining parts of the field equations.
Consequently, for such proposed solutions one finds
$$\eqalignno{
&G^r_r-\kappa T^r_r=-2a^{-1}\dot a\d&(6a)\cr
&\nabla^2\phi-\p V/\p\phi=-\dot\phi\d.&(6b)\cr}$$
Evidently the Einstein-Klein-Gordon equations are not satisfied.
Instead, the proposed solutions describe Dirac $\d$ sources at the junction.
The $\d$ term in the Einstein equation describes a spatial layer of matter 
with pressure but no energy or momentum,
according to the physical interpretation of Einstein's theory,
while the $\d$ term in the Klein-Gordon equation 
may be interpreted as a step potential [4].

Returning to the issue raised by the original article [1]:
the proposed solutions violate conservation 
because they violate the field equations.
\np\ni
{\sl III. Hellaby \& Dray's criticisms}

Hellaby \& Dray's strategy is to somehow escape the above facts 
and justify their proposed solutions.
Their main tactic is to argue that 
these facts depend on a particular narrow view 
of field equations and solutions,
whereas their more liberal philosophy allows them an alternative approach,
in which their proposed solutions magically satisfy the field equations 
after all.
Their reply [3] is titled as a ``comparison of approaches''.
But their comparisons are false, as follows.

(i).  Hellaby \& Dray repeatedly claim that they obtain different results
because I use ``imaginary time'' and they use real time.
I have never used imaginary time.
In the calculation of [2], repeated above, $t$ and $\tau$ are real, 
as are all other quantities.
In all my papers, all time coordinates are real.
The only occasions I have mentioned imaginary time 
have been to emphasize that I am using real time.

(ii).  Hellaby \& Dray also repeatedly claim that they obtain different results
because their approach is ``purely classical'' and mine is not.
The calculation of [2], repeated above, is purely classical and obviously so:
only the classical field equations are considered.
All my work on signature change is purely classical
except my most recent preprint [5],
which considers quantum theory via path integrals.

(iii).  Hellaby \& Dray repeatedly claim that the calculation of [2], 
repeated above, 
is based on a ``modified form'' of the field equations,
thus allowing them their own modifications.
The calculation of [2], repeated above, 
is evidently of the usual Einstein-Klein-Gordon equations
with no modification.

(iv).  Hellaby \& Dray repeatedly claim that 
there is no rigourous derivation of distributional field equations.
(Since this presumably includes their own attempted derivations,
it is a noteworthy admission).
The calculation of [2], repeated above, is such a derivation.

Hellaby \& Dray's abstract, introduction and conclusion [3] 
consist almost entirely of these absurd criticisms, presented as comparisons.
The publication of such obviously false accusations in a refereed journal 
is unacceptable.
\ms\ni
{\sl IV. Hellaby \& Dray's attempted justification}

Hellaby \& Dray's other main tactic for rescuing their proposed solutions
is to pretend that they were solving different equations.
This hilarious bluff rests on the fact that their proposed solutions
fail to satisfy the field equations only at the junction, 
by the $\d$ terms (in unit normal coordinates).
Their idea is to delete the $\d$ term from the Einstein equations
and ``postulate'' the resulting equations 
as an ``alternative form'' of the Einstein equations,
just as good as anyone else's.
Noting that Hellaby \& Dray use $\epsilon=-\e$,
their desired equations (33)--(34) are immediately recognisable 
as the above Einstein equations (4a)--(4b) with the $\d$ term missing.
They precede these equations with a convoluted argument
supposedly deriving them from their original Einstein equations (2)--(3).
Since their argument is carefully constructed to fool the reader,
it may be useful to identify the various invalid steps 
and correct the identity on which their derivation hinges, as follows.

In (26)--(28), they have set $[a]\d$, $[\phi]\d$ and $[a']\d$ equal to zero,
whereas substitution of these expressions into (2)--(3) requires 
multiplication of such distributions by themselves and division by zero,
neither of which is well defined.
For instance, there should be a term $[a]^2\d^2/a^2N$ 
involving the ill defined square of $\d$.
There should also be a term $[a']\d/aN$ which is of the form $0/0$,
since the authors are assuming here that $N\to0$ as $t\to0$.
(Their (31) also involves division by zero).
Such terms have been lost in their calculation.
One cannot simply set $0/0$ equal to $0$;
one must show that the numerator approaches zero faster than the denominator.
For instance, the term $[a']\d/N$ is zero if $a'=O(N)$ as $N\to0$,
which gives the junction condition (5a) above.

This has led to Hellaby \& Dray's incorrect equation (29):
$$\epsilon\ddot a={N'a'\over{2N^2}}-{a''\over{N}}.$$
They use this as an identity relating second derivatives 
with respect to the proper time $\tau$ 
and the original unspecified time coordinate $t$,
by replacing the right-hand side with the left in the Einstein equations.
The proposed identity may be checked simply by substituting $t=\tau$.
This corresponds to $N\epsilon=-1$,
according to their definition (4):  $d\tau=\sqrt{-\epsilon N}dt$. 
Formally substituting this choice of $N$ and noting $\dot\epsilon=-2\d$, 
(29) reads
$$\epsilon\ddot a=-\d\dot a+\epsilon\ddot a.$$
This is obviously not self-consistent.
It clearly illustrates how Hellaby \& Dray have lost the $\d$ term 
in the Einstein equations.
It appears to be a deliberate ploy to fool the reader by sleight of hand;
the authors have known about the $\d$ term for years,
but tried to lose it during an unnecessary detour.
Direct substitution of $t=\tau$ would have worked.

The choice $N\epsilon=-1$ is formally equivalent to 
the choice $\lambda=\e$ used above,
recalling $N=\l^{-1}$ and $\epsilon=-\e$.
Direct substitution into the d'Alembertian $\tilde\nabla^2$ 
of the normal space yields
$$\tilde\nabla^2f=-\l f''-\half\l'f'=-\e\ddot f-\d\dot f.\eqno(7)$$
This identity correctly relates second derivatives 
with respect to a general time coordinate $t$ and the proper time $\tau$.
The $\d$ term cannot be argued away.

This latest approach of Hellaby \& Dray 
also differs from that of their original paper [1] 
and their previous papers with different co-authors.
Originally, the various authors simply solved the field equations 
either side of the junction and joined the solutions together 
according to obvious continuity conditions,
without checking whether this ensured that 
the field equations were satisfied at the junction.
This turns out not to be so, but this fact has made little headway 
against the authors' deep attachment to their proposed solutions.
By now, a recognisable cycle has emerged:
the authors claim to have justified their proposed solutions,
I point out where their argument is invalid,
then they write another paper denying the error 
and claiming another justification by a different method, and so on.
This cycle has repeated several times already
and seems set to continue until editors and referees wake up.
It is a ludicrous way to deal with having solved an equation incorrectly.

Rather than give up their proposed solutions, 
Hellaby \& Dray's latest form of self-justification is to change the equations.
They apparently believe that 
they can remove the $\d$ terms from the field equations 
(in unit normal coordinates) by some sufficiently complicated manipulation.
Not surprisingly, their attempts to do so are invalid.
Hellaby \& Dray seem to realise this,
since they ``postulate'' their desired equations (33)--(34)
immediately after supposedly deriving them.
The fact remains that these equations 
differ from the Einstein-Klein-Gordon equations, by well defined terms.

In the remainder of their reply [3], Hellaby \& Dray avoid 
attempting to write down the full field equations.
Instead they discuss ``imaginary versus real'' time 
and ``choice'' of junction conditions.
These diversions and their argument that 
``the difference in results is entirely explained 
by the difference in philosophy'' hardly deserve further comment.
\ms\ni
{\sl V. Conclusion}

It has been shown that Hellaby \& Dray's attempted derivation 
of ``alternative'' field equations is invalid,
that their main criticisms of my previous comment [2] are false,
and that their proposed solutions fail to satisfy the field equations 
by certain well defined Dirac $\d$ terms 
with a standard physical interpretation as sources at the junction.

The situation has become ridiculous:
the correction of a straightforward error has required
a comment on a reply to a comment on the original paper,
which itself was written after the error (in previous papers)
had already been identified in published work.
Hellaby \& Dray and their various co-authors are simply denying the facts,
thereby creating an entirely fruitless controversy.
Some might accept such behaviour as psychologically inevitable,
but it is not inevitable that scientific journals publish the resulting drivel.
\bs
\begingroup
\parindent=0pt\everypar={\global\hangindent=20pt\hangafter=1}\par
{\bf References}\par
\ref{[1] Hellaby C \& Dray T 1994}\PR{D49}{5096}
\ref{[2] Hayward S A 1995}\PR{D52}{7331}
\ref{[3] Hellaby C \& Dray T 1995}\PR{D52}{7333}
\refb{[4] Hayward S A 1995}
{Signature change at material layers and step potentials}{(gr-qc/9509052)}
\refb{[5] Hayward S A 1995}
{Complex lapse, complex action and path integrals}{(gr-qc/9511007)}
\endgroup
\bye